\documentclass[twoside,reqno]{bjp}

\usepackage{amssymb,amsmath}
\usepackage{times}
\newcommand{\ie}{{\em i.e.~}}

\newcommand{\be}{\begin{equation}}
\newcommand{\ee}{\end{equation}}
\newcommand{\ba}{\begin{eqnarray}}
\newcommand{\ea}{\end{eqnarray}}

\begin{document}

 \sloppy \raggedbottom

 \setcounter{page}{1}



\title{Correlation functions of conserved currents in four dimensional conformal field theory with higher spin symmetry\thanks{Invited talk at the 
  Second Bulgarian National Congress in Physics, Sofia, September 2013}}

\runningheads{Correlation functions of conserved currents in  CFT with HS symmetry}{Yassen S. Stanev}

\begin{start}
\author{Yassen S. Stanev}{}

\address{INFN Sezione di Roma Tor Vergata, 00133 Rome, Italy}{}


\begin{Abstract}
We report some recent progress in the computation of the $n$-point correlation functions of conserved currents in a class of four dimensional conformal field theories with higher spin symmetry. Global conformal invariance leads to very strong constraints on both the general form and the singularity structure of the correlation functions of conserved currents. Namely, all these functions have to be rational functions with at most double pole singularities. We show that this implies that the 4-, 5- and 6-point correlation functions of the (symmetric, conserved and traceless) stress-energy tensor are linear combinations of the three free field expressions. Hence, in four dimensions, any globally conformal invariant theory is free.
\end{Abstract}

\end{start}

\section[]{Introduction. Summary of the results}

In the last decade an approach, particularly well suited for the study of  Conformal invariant quantum Field Theory (CFT) in four dimensions, has been developed. 
It is based on the notion of Global Conformal Invariance (GCI) \cite{NT}
and makes use of bilocal and biharmonic conformal fields \cite{biharm1}. 
In a GCI theory all the bosonic fields have integer scale dimension, all the $n$-point correlation functions are rational functions,
and there are always infinitely many Higher Spin (HS) conserved currents. 
Moreover, if in the theory there is a hermitian scalar field $\Phi_2$ of scale dimension equal to two, then all the functions can be realized in terms of free scalar fields \cite{Rational1,Rational2}.
There has been also an important progress in understanding the general structure of theories with HS symmetry. In \cite{Boulanger:2013zza} it was shown that introducing a single HS conserved current always leads to infinitely many conserved currents. All the 3-point functions of the HS conserved currents have been found in \cite{GenCFT}, while all the $n$-point functions of the free HS conserved currents were derived in \cite{Gelfond:2013xt}. 
In three dimensions it has been proven 
that all the correlation functions of the observables  in any CFT  
with HS symmetry reduce to the ones in a free field theory  \cite{MZ}. 
This can be viewed as an extension of the Coleman-Mandula theorem \cite{Colem-M} to the CFT case.

In this paper we extend the Coleman-Mandula theorem to a class of four dimensional CFTs with HS symmetry. 
To be precise, we consider unitary GCI theories with all the standard properties in four dimensional Minkowski space (existence of a symmetric, traceless and conserved stress-energy tensor $\Theta_{\mu \nu}(x)$, Operator Product Expansion (OPE), cluster decomposition, etc.).  
Starting from the stress-energy tensor OPE we construct an auxiliary conformal scalar bi-field $V(x_1,x_2)$, which biharmonic part $V_2(x_1,x_2)$ projects on the contributions of the HS conserved currents in the stress-energy tensor OPE.
Thus $V_2$ captures the important part of the OPE directly related to the symmetries of the theory. The stress energy tensor $\Theta_{\mu \nu}(x)$ can be extracted from $V_2$ by an appropriate limit.
In other words $V_2$ interpolates in the fusion defined by the OPE
\be 
\Theta_{\mu \nu} \;  \Theta_{\rho \tau} \ \rightarrow \ V_2 \ \rightarrow \ \Theta_{\alpha \beta}  \ .
\label{fusion}
\ee
Hence, one can derive the $n$-point functions of the stress-energy tensor ${\cal G}_{n}$ from the $2n$-point functions of $n$ biharmonic fields $V_2$, which as we shall argue are easier to compute. Note that the correlation functions of the stress-energy tensor are a coupled system and have to be considered together rather than separately, \ie a given  $n$-point function ${\cal G}_{n}$ is allowed only if there exists an (infinite) tower of $n+k$-point functions ${\cal G}_{n+k}$ which can be reduced to ${\cal G}_{n}$  by the fusion defined in (\ref{fusion}). Combining this with the constraints implied by conformal invariance and HS symmetry, it follows that in any GCI theory all the $n$-point functions of the stress-energy tensor will have at most double pole singularities in all arguments (a property typical for the free field theories).
Finally we compute the 4-, 5- and 6-point correlation functions of the  stress-energy tensor and find that they are linear combinations of the three free field expressions (scalar, fermion and Maxwell field), see Eq.(\ref{T6}). 
It follows that any unitary GCI theory is free.

\section{Biharmonic field construction}

The stress-energy tensor $\Theta_{\mu \nu}(x)$ plays a central role in any CFT since it generates the conformal transformations. It is symmetric, traceless and conserved and 
has scale dimension $\Delta_\Theta=4$. In a theory with HS symmetry, there are also (infinitely many) 
HS conserved currents, namely symmetric traceless conserved tensors $J_r^{(\mu_1 \dots \mu_r)}(x)$ of rank $r$ and scale dimension $\Delta_r=r+2$, transforming in the Lorentz representation $(r/2,r/2)$.  For a symmetric tensor, the quantity $\Delta_r-r$ is called the twist, so all the conserved currents $J_r(x)$ have twist two. The $r=0$ case is a scalar field $\Phi_2$ of scale dimension equal to two, the $r=1$ field is the usual dimension three conserved current $J_{\mu}(x)$, while the  stress-energy tensor $\Theta_{\mu \nu}(x)$ is (one of) the $r=2$ fields. 

Consider the truncated $n$-point Wightman functions  of the stress-energy tensor  
\be
{\cal G}_n(x_1,x_2,\dots,x_n) = \langle \Theta^{\mu_1 \nu_1}(x_1) \Theta^{\mu_2 \nu_2}(x_2) \dots \Theta^{\mu_{n} \nu_{n}}(x_{n}) \rangle \vert_{\rm truncated}
\, ,
\label{defGn}
\ee
where, as usual, truncated means that the terms which are products of lower point functions are subtracted.
${\cal G}_n$ is totally symmetric under the permutation of any two arguments (accompanied by the permutation of the respective Lorentz indices).
The properties of the stress-energy tensor $\Theta_{\mu \nu}(x)$ imply that ${\cal G}_n$ is a parity even, symmetric and traceless in each pair of indices  $(\mu_i,\nu_i)$ and conserved in all its arguments function. 
The leading short distance singularities of the 
functions Eq.(\ref{defGn}) are determined by the OPE, which is rather complicated \cite{Me88}, so we shall briefly review only some relevant for our discussion properties. The general form of the OPE is
\be
 :\Theta_{\mu \nu}(x_1) \, \Theta_{\rho \tau}(x_2): \ = \   \sum_{R}
\  {\cal C}^{R}_{\mu \nu \rho \tau} (x_{12},\partial_{x_2}) \ {\cal O}_R(x_2) \, ,
\label{TTOPE}
\ee
where $x_{12}=x_1-x_2$ and
 the label $R =(j_1,j_2;\Delta)$ defines the representation of the conformal group  to which belongs the field ${\cal O}_R$.
Combining scale invariance and unitarity of the conformal representations \cite{Mack} leads to upper bounds for the order of the poles 
in $x_{12}^2$ of the coefficient functions  ${\cal C}^{R}$. The explicit expressions for these pole bounds are given in \cite{Free4DHS}.

Given the OPE of  two stress-energy tensors $\Theta_{\mu \nu}$, Eq.(\ref{TTOPE}), one can construct the auxiliary  
bi-field $V(x_1,x_2)$  
\ba
V(x_1,x_2) &=&  x_{12}^2  \, \left( x_{12}^\mu x_{12}^\nu x_{12}^\rho x_{12}^\tau 
- x_{12}^2 \, x_{12}^\mu x_{12}^\rho \, \eta^{\nu \tau} + \frac{x_{12}^4} {4}  \, \eta^{\mu \rho} \, \eta^{\nu \tau} 
 \right)  \nonumber  \\
&\times& \, :\Theta_{\mu \nu}(x_1) \, \Theta_{\rho \tau}(x_2):  \ . 
\label{defV}
\ea
The bi-field  $V(x_1,x_2)$ has a number of interesting properties:
It transforms as a scalar conformal bi-field of weights $(1,1)$ in $x_1$ and $ x_2$ respectively.
It is symmetric under the exchange of
$x_1$ and $x_2$, hence it receives contributions from only the even rank symmetric tensors in the OPE $\Theta_{\mu \nu}(x_1) \, \Theta_{\rho \tau}(x_2)$.
It is finite in the limit $x_{12} \rightarrow 0$. 
The non-zero limit of $V(x_1,x_2)$ for coincident arguments is in one-to-one correspondence with the presence of scalar field $\Phi_2$ of scale dimension two.  Hence, for the search of interacting theories, without loss of generality\footnote{The case of a GCI theory 
generated by the scalar field $\Phi_2$ has been analyzed in detail in \cite{Rational1} and it has been shown that it reduces to a 
theory of free scalars.} we can set $V(x_1,x_1) \ = \ 0 $.
In the light-cone limit, when $x_{12}^2 \rightarrow 0$, only the even rank twist=2 conserved currents $J_r$ contribute to $V(x_1,x_2)$. 
Even if several different rank two symmetric traceless conserved
tensors contribute to the $\Theta \Theta$ OPE, only the stress-energy tensor
$\Theta_{\mu \nu}$ contributes to  $V(x_1,x_2)$. 
The restriction of $V$ to only the twist=2 contributions
in the stress-energy tensor OPE 
\be
V_2(x_1,x_2) = V(x_1,x_2) \vert_{{\rm twist}=2} = \sum_r {\cal C}_r(x_{12},\partial_{x_2}) \,  J_r(x_2) \, ,
\label{V2}
\ee
is biharmonic $\Box_{x_1} V_2(x_1,x_2) = 0 = 
\Box_{x_2}  V_2(x_1,x_2)$.

Since by assumption the scalar field $\Phi_2$ is absent in $V$, the stress-energy tensor $\Theta_{\mu \nu}$ is the leading (for small $x_{12}$) contribution in $V(x_1,x_2)$ and can be expressed  as 
\be
\Theta^{\mu \nu}(x_1) = 
\left( \partial_{x_{12}}^{\mu} \partial_{x_{12}}^{\nu} 
- \frac{\eta^{\mu \nu}}{4} \, \Box_{x_{12}} \right) V(x_1,x_2)\vert_{x_{12}=0}  \, . 
\label{ThetafromV}
\ee
Note that we can replace $V(x_1,x_2)$ with $V_2(x_1,x_2)$ in this equation.
To summarize, Eqs.(\ref{defV}),(\ref{ThetafromV}) define the fusion procedure
Eq.(\ref{fusion}). 
Thus, one possible way to compute all the possible conformal invariant $n$-point functions of the stress-energy tensor $\Theta_{\mu \nu}$ in any CFT is to classify all the $2n$ point functions of $V_2$, harmonic in all the arguments and then use  Eq.(\ref{ThetafromV}) to derive the respective functions of the stress-energy tensor. 

Let us denote the $2n$-point Wightman function of $n$  bi-fields $V$  by ${\cal W}(2n)$ and the $2n$-point 
Wightman function of $n$  biharmonic fields $V_2$  by  ${\cal W}_2(2n)$.
Both functions are scalar conformal invariant functions of conformal weights in all the arguments equal to one. Hence they both will depend only on\footnote{With the $+{\rm i} 0  x_{jk}^0$ prescription in the denominators.}
$\{x_{jk}^2\}$.
However there is an important difference. 
Since  ${\cal W}(2n)$ is obtained from the rational function  ${\cal G}_{2n}$
by multiplication with $x_{ij}$ and contracting the Lorentz indices,  
 it is also a rational function,  
while the function ${\cal W}_2(2n)$, harmonic in all its arguments, is not related in a simple way to ${\cal G}_{2n}$ and apriori is not rational. Note however, that the restrictions of both the functions ${\cal W}(2n)$ and ${\cal W}_2(2n)$ for
$x_{12}^2=x_{34}^2= \dots = x_{2n-1 \, 2n}^2=0$ coincide, since in this limit one projects out all the contributions of the higher twist fields (as well as some of the higher derivatives of the twist=2 fields) 
\be
{\cal W}_0(2n) = {\cal W}(2n)\vert_{\{x_{2k-1 \, 2k}^2=0\}} 
= {\cal W}_2(2n)\vert_{\{x_{2k-1 \, 2k}^2=0\}} \ .
\label{W0}
\ee
Hence, on the one hand ${\cal W}_0(2n)$ is a rational function, on the other hand it can be completed to a harmonic in all its arguments function. 
Note that, while any rational function can be 
completed to a harmonic in one of its arguments function \cite{BargTod}, the 
requirement that there exists a completion harmonic in all arguments is highly non-trivial. In \cite{Free4DHS}, using results of \cite{biharm1}, it has been proven that the rational function ${\cal W}_0(2n)$
must have the so called "single pole" singularity structure, 
\ie in the denominator of each term of ${\cal W}_0(2n)$ each coordinate $x_i$ 
appears only once. 
Then the unique multi-harmonic completion ${\cal W}_2(2n)$ of ${\cal W}_0(2n)$ is again a rational function with single pole singularities. 

Given the function ${\cal W}_2(2n)$,
one can use Eq.(\ref{ThetafromV}) to compute the $n$-point function of 
the stress-energy tensor ${\cal G}_{n}$ 
\be
{\cal G}_n(x_1,x_3,\dots,x_{2n-1}) 
 \  = \ \prod_{k=1}^{n}{\cal D}_{2k-1} {\cal W}_2(2n) \vert_{\{x_{2k} = x_{2k-1}\}}
\, ,
\label{TfromW0}
\ee
where ${\cal D}_{2k-1}$ are obtained by the substitution $1 \rightarrow 2k-1$, $2 \rightarrow 2k$ from
\be 
{\cal D}_{1}  = \left( \partial_{x_{12}}^{\mu_{1}} \partial_{x_{12}}^{\nu_{1}} 
- \frac{ \eta^{\mu_{1} \nu_{1}}}{4} \, \Box_{x_{12}} \right) \,  . \nonumber
\label{CalD1}
\ee
Since  ${\cal W}_2(2n)$ has only single pole 
singularities, all the $n$-point functions of the stress-energy tensor ${\cal G}_{n}$ will have at most double pole singularities. This is trivially satisfied for all the free field theories. 
Indeed for free fields, the stress-energy tensor is 
a bilinear combination of the fundamental fields and all its functions can be expressed in terms of the 2-point functions of the fundamental fields by Wick theorem, leading to double pole structure. As we shall now show the double pole property is satisfied only in the free field theories.

\section{Computing the correlation functions}
\label{sec:computation}

Eq.(\ref{TfromW0}) allows to express  the $n$-point function of the stress energy tensor ${\cal G}_{n}$ in terms of ${\cal W}_2(2n)$, the multi-harmonic completion of the restriction ${\cal W}_0(2n)$ of the function ${\cal W}(2n)$
obtained from the $2n$-point function of the stress energy tensor ${\cal G}_{2n}$. The results of the previous section imply that for any given ${\cal G}_{2n}$, ${\cal W}_2(2n)$ is a function of $2n$ variables $x_1,\dots,x_{2n}$, with all the following properties:
scalar,  
symmetric under the permutation of  $x_{2k-1}$ and $x_{2k}$ for any $k$,
symmetric under the permutation of any two pairs  $(x_{2j-1},x_{2j})$ and $(x_{2k-1},x_{2k})$, 
conformal invariant of conformal weight one in all its arguments,
harmonic in all its arguments,
rational,
single pole  in all its arguments with
maximal order of all the poles (in $x_{ij}^2$) less or equal to five, 
vanishing in the limit $x_{2k} \rightarrow x_{2k-1}$ for any $k$. 
Since all the properties are linear,
for any fixed value of $n$ the general function $F$ with all the above properties will be a finite linear combination $F= \sum_k c_k F_k$,  of functions $F_k$ which again have all the properties, with  arbitrary coefficients $c_k$. For generic 
choice of these coefficients the function  $F$ may not correspond to a restriction of any $2n$-point function of the stress-energy tensor ${\cal G}_{2n}$. However the restrictions of all possible $2n$-point functions ${\cal G}_{2n}$ will certainly correspond to some choice of $c_k$, hence also 
all possible $n$-point functions ${\cal G}_{n}$ can be obtained 
by applying the differential operator in the r.h.s of  Eq.(\ref{TfromW0}) to  $F$. 
For  $n$=3,4,5 and 6 we derived the general function $F(x_1,\dots,x_{2n})$ with all the above properties and computed 
\be
f(x_1,x_3,\dots,x_{2n-1}) =  \prod_{k=1}^{n}{\cal D}_{2k-1}  F(x_1,\dots ,x_{2n}) \vert_{\{x_{2k} = x_{2k-1}\}}
\, ,
\label{TfromF}
\ee
then we imposed on $f$ the pole bounds coming from the OPE and the cluster decomposition (in order to exclude for $n=6$ the products of 3-point functions). 
The result of this rather challenging, since for $n=6$ the respective 12-point function had several billions of terms, calculation is: in all cases $f$ contains only the three free field $n$-point functions.

In other words, for any $n \leq 6$ the general GCI 
$n$-point function of the stress-energy tensor, defined in Eq.(\ref{defGn}), is a linear combination of the three free field expressions (free scalar, free fermion and free Maxwell field)
\be 
 {\cal G}_{n} = c_{\varphi}\langle\Theta\dots\Theta\rangle_{\varphi}
+c_{\psi}\langle\Theta\dots\Theta\rangle_{\psi}
+c_{F}\langle\Theta\dots\Theta\rangle_{F} \, ,
\label{T6}
\ee
where the coefficients $c_{\varphi}$, $c_{\psi}$ and $c_{F}$ 
are the same for all values of $n$.
The spectrum of the fields ${\cal O}_R$ and the coefficient functions 
${\cal C}^R$,  appearing in the OPE Eq.(\ref{TTOPE}) are determined completely
already from the 4-point function.  
The 6-point function, by performing a triple OPE, implies that all the 3-point functions of all the operators which appear in the OPE of the stress-energy tensor are equal to the three free field ones. This is highly non-trivial, since in general there are many more conformal invariant structures for the 3-point functions \cite{GenCFT}, and essentially implies that all the GCI theories reduce to the three free field ones.

Let us also note that in \cite{Alba:2013yda} the same result was derived under the (very strong) assumption that in the CFT with HS symmetry the stress-energy tensor is the unique rank two conserved current.

\section*{Acknowledgments}

It is a pleasure to thank the organizers of the Second Bulgarian National Congress in Physics, and in particular Vladimir Dobrev.
This work was supported in part by MIUR-PRIN contract 2009-KHZKRX-005.

\end{document}